\documentclass[epj,nopacs]{svjour}
\usepackage{epsfig,array}
\usepackage{multirow,color}
\usepackage{graphicx,float}
\newcommand{\be}{\begin{eqnarray}}
\newcommand{\ee}{\end{eqnarray}}

\newcommand{\er}{$\pm$}

\newcommand{\bea}{\begin{eqnarray}}
\newcommand{\eea}{\end{eqnarray}}

\newcommand{\beq}{\begin{equation}}
\newcommand{\eeq}{\end{equation}}

\def\fun#1#2{\lower3.6pt\vbox{\baselineskip0pt\lineskip.9pt
\ialign{$\mathsurround=0pt#1\hfil##\hfil$\crcr#2\crcr\sim\crcr}}}

\begin{document}

\title{\boldmath Interference phenomena in the $J^P=1/2^-$-wave in $\eta$ photoproduction}
\titlerunning{Interference phenomena in the $J^P=1/2^-$-wave in $\eta$ photoproduction}
\author{A.V.~Anisovich$\,^{1,2}$, E.~Klempt$\,^1$, B. Krusche$\,^3$,
  V.A.~Nikonov$\,^{1,2}$,  A.V.~Sarantsev$\,^{1,2}$, U.~Thoma$\,^1$, D. Werthm\"uller$\,^3$
\thanks{Present address: School of Physics and Astronomy, University of Glasgow, Glasgow G12 8QQ, United Kingdom}\\ }
\authorrunning{A.V.~Anisovich \it et al.}
\institute{$^1\,$Helmholtz-Institut f\"ur Strahlen- und Kernphysik,
Universit\"at Bonn, Germany\\
$^2\,$Petersburg Nuclear Physics Institute,  Gatchina, Russia\\
$^3\,$Departement f\"ur Physik, Universit\"at Basel, Switzerland
}
\date{Received: \today / Revised version:}

\abstract{The recent precise experimental results for the photoproduction
of $\eta$-mesons off the neutron measured with the Crystal Ball/TAPS calorimeter
at the MAMI accelerator have been investigated in detail in the framework of the 
Bonn-Gatchina coupled channel model. The main result is that the narrow structure 
observed in the excitation function of $\gamma n \rightarrow n\eta$ can be 
reproduced fully with a particular interference pattern in the $J^P=1/2^-$ partial
wave. Introduction of the narrow resonance $N(1685)$ with the properties reported in earlier publications deteriorates the quality of the fit.
}
\maketitle

\section{Introduction}

So far photoproduction of mesons off the neutron has been much less
investigated than the corresponding reactions off the free proton.
The reasons are the obvious difficulties related to measurements
using nucleons bound in nuclei (in most cases neutrons bound in the
deuteron) as targets. There are not only the technical complications
arising from the necessity to detect the recoil neutrons but also
the difficulties in the interpretation of the results which are
effected by nuclear Fermi motion and Final State Interaction (FSI)
effects. Nevertheless, such reactions are important because they reveal
the isospin structure of the electromagnetic excitation currents.
The coupling of isospin $I=3/2$ $\Delta$ resonances to $\gamma N$
is identical for protons and neutrons, but the $\gamma N\rightarrow N^{\star}$
couplings are isospin dependent. During the last few years quite some
progress has been made for this branch of the photonuclear experimental
program \cite{Krusche:2011} and first results have been reported
for several reaction channels. The measurements of $\eta$ photoproduction
off the neutron have attracted particular interest, because around 1 GeV
of incident photon energy ($W\approx$1680~MeV) a narrow structure was
observed in the excitation function \cite{Kuznetsov:2006kt,Jaegle:2008ux,Jaegle:2011sw}. These observations are listed by the Particle Data Group \cite{Beringer:1900zz} as one-star nucleon resonance $N(1685)$. Remarkably, such a 
structure had been predicted by soliton models in the context of the
conjectured baryon antidecuplet of pentaquarks. The nonstrange 
member of the multiplet with spin-parity $J^P=1/2^+$ \cite{Diakonov:1997mm} should be electromagnetically
excited more strongly on the neutron, should have a large decay branching
ratio to $N\eta$, an invariant mass around 1.7\,GeV, and a width
of a few tens of MeV \cite{Diakonov:1997mm,Polyakov:2003dx,N_1680,Arndt:2003ga};
all properties that are phenomenologically exhibited by the observed structure.
If one treats this structure as a single isolated resonance, a mass of $1670\pm5$\,MeV and a width  
of $\Gamma\approx30$\,MeV are determined. Assuming a constant
angular distribution ($J^P=1/2^+$  or $1/2^-$ ) and ignoring possible interference effects,
the electromagnetic coupling strength is determined to
$A_{1/2}\cdot\sqrt{b_{\eta}}\approx 12\times 10^{-3}$\,GeV$^{-1/2}$ \cite{Jaegle:2011sw} or,
respectively, to
$(12.3\pm0.8)\times 10^{-3}$\,GeV$^{-1/2}$
\cite{Werthmuller:2014}. The radiative width derived by Azimov {\it et al.}~\cite{Azimov:2005}
using the GRAAL data \cite{Kuznetsov:2006kt} corresponds to $A_{1/2}\cdot\sqrt{b_{\eta}}=15\times 10^{-3}$\,GeV$^{-1/2}$. The experimental values for the properties of $N(1685)$ 
are all in the predicted range for the nonstrange partner
of the $\Theta^+$, an exotic baryon which was ``discovered"  in 2003. Shortly after, the evidence
for its existence faded away in a number of precision experiments \cite{Burkert:2005ft,Hicks:2012zz,Liu:2014yva} but evidence is reported in several more recent experiments \cite{Nakano:2008ee}, \cite{Amaryan:2011qc} (see, however, \cite{Anghinolfi:2012np}), and \cite{Barmin:2013lva}.

In contrast to
the history of the exotic $\Theta^+$ pentaquark, the statistical significance
of the structure observed in $\gamma n\rightarrow n\eta$ is undisputable. All
experiments that searched for this structure came out with positive results,
and the most recent measurements at the MAMI accelerator with deuterium
\cite{Werthmuller:2014,Werthmuller:2013rba} and also $^3$He targets
\cite{Werthmuller:2013rba,Witthauer:2013} established it beyond any doubts.
Due to the full kinematic reconstruction of the $\eta$ - neutron
final state, effects from nuclear Fermi motion - smearing out narrow
structures - were removed so that a better estimate of the width
of the structure became possible~\cite{Werthmuller:2014,Werthmuller:2013rba}. Monte Carlo simulations showed that the observed width of 50$\pm$10\,MeV (the natural width folded with the experimental resolution) corresponds to a natural width of
only $\approx$30\,MeV. Such a narrow width would be very unusual for a normal three-quark
nucleon resonance with a mass of W$\approx$1680\,MeV. There is a possibly correlated,
although very week effect, in $\gamma p\to\eta p$ \cite{McNicoll:2010qk}
where a narrow dip is observed at the same incident photon energy.
An observation has also been reported for Compton scattering $\gamma n\to\gamma n$ \cite{Kuznetsov:2010as}. A measurement of the beam asymmetry $\Sigma$ for Compton scattering off  protons suggested even two narrow structures with masses near 1680 and 1.720\,MeV \cite{Kuznetsov:2015nla}.

We have thus the akward situation that a resonance was predicted at about 1680\,MeV in the soliton model, and a bump-like structure with exactly the right properties was found in experiments.
However, there are serious doubts that this bump is related to the predicted
pentaquark state. Here, one should not forget that not each bump observed in some
excitation function is evidence for a resonance. There can be other effects, for
example threshold cusps. In fact, it has already been tried to model this bump
with various approaches. Apart from intrinsically narrow states
\cite{Arndt:2003ga,Choi:2006,Fix:2007,Shrestha:2012,Anisovich:2008wd},
different coupled-channel and interference effects of known nucleon resonances
 have been discussed in the literature. The Gie\ss en group claimed that the narrow peak in the $\eta$ photoproduction on the neutron can be explained as 
$N(1650)1/2^-$ and $N(1710)1/2^+$ coupled-channel effect~\cite{Shklyar:2006xw}; Shyam and Scholten  use interference effects between the $N(1650)1/2^-$, $N(1710)1/2^+$, and $N(1720)3/2^+$ resonances  to describe the peak~\cite{Shyam:2008fr}, D\"oring and Nakayama ascribe the peak to effects from
strangeness threshold openings \cite{Doering:2009qr}. The Bonn-Gatchina group demonstrated that the narrow peak can be explained naturally by interference effects in the $J^P=1/2^-$ wave \cite{Anisovich:2008wd}, a conclusion which was confirmed independently - even though three years later - by Zhong and Zhao~\cite{Zhong:2011ti}. 

Here, we come back to the idea first put forward in \cite{Anisovich:2008wd} where two different scenarios for the bump structure in the CBELSA $\gamma n\rightarrow n\eta$ data
\cite{Jaegle:2008ux} where discussed. In the first scenario, the two resonances with spin-parity $J^P=1/2^-$ - $N(1535)1/2^-$ and $N(1650)1/2^-$ - and the interference between them was studied; this ansatz gave a good fit to the data. In the second scenario, a narrow resonance at 1685\,MeV and with photo-coupling and $N\eta$ decay branching ratio as predicted in \cite{Diakonov:1997mm,Polyakov:2003dx,N_1680,Azimov:2005} was introduced; this model gave a fit of equivalent quality. Thus the existence of $N(1685)$ was not supported but could not be ruled out. This situation has now completely changed with the new, very precise data measured by the Crystal Ball/TAPS experiment at MAMI \cite{Werthmuller:2014,Werthmuller:2013rba}. In this paper we compare fits to the new data with the two scenarios discussed above and find a much better description without introduction of the $N(1685)$ resonance, which makes the interpretation of the bump-like structure as resonance improbable.

In the experimental publications describing the new data \cite{Werthmuller:2014,Werthmuller:2013rba},
the same strongly simplified model as in \cite{Jaegle:2008ux,Jaegle:2011sw}
was used. The model consists of a superposition of three Breit-Wigner functions without interference terms. the Breit-Wigner functions represent
i) the narrow structure, ii) the $N(1535)1/2^-$ standing for the full $J^P=1/2^-$ partial wave, and iii) the background contributions. The model served to extract phenomenological estimates for position and width of the
observed bump. 

It is obvious that a much refined analysis of the data is necessary and this is
possible within the Bonn-Gatchina coupled channel approach, which was recently
updated for reactions off neutrons \cite{Anisovich:2013jya}. This paper discusses
in detail new fits to the data within different scenarios and their implication to
the nature of the observed structure. A comment-like short version of part
of this analysis is already available on the arXiv \cite{Anisovich:2014hga} .

The paper is organized as follows. In section~\ref{exp} we summarize the experimental
data used in this paper.  Subsequently (section~\ref{sim}), we present some simulations;
the aim is to introduce to the reader the patterns which may emerge from known input
into a partial wave analysis. In section~\ref{pwa} we present our fits to the
data imposing or not imposing the presence of a narrow resonance in the $J^P=1/2^+$
wave. A short summary is given at the end.

\section{\label{exp}Data used in this analysis}
\begin{figure}[pt]
\begin{center}
\epsfig{file=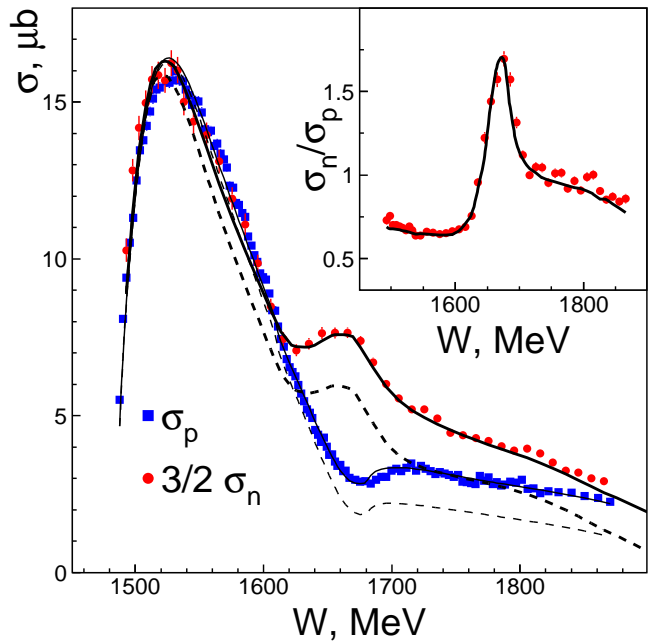,width=0.48\textwidth}
\vspace{-2mm}\end{center}
\caption{\label{tot} (color online) The total cross section for $\gamma n\to \eta n$, $\gamma p\to \eta p$, and their ratio as functions of the $\eta N$ invariant mass. The solid curves represent our final fits, dashed curves the $J^P=1/2^-$ contributions.
}
\end{figure}

The main new data used in this paper  are the $\gamma n\to \eta
n$ differential cross section from MAMI
\cite{Werthmuller:2014,Werthmuller:2013rba}. The data -- shown in
Fig.~\ref{tot} -- were taken with a deuteron target but with  full
event reconstruction. Therefore these data do not suffer from the
Fermi motion which usually smears out narrow structures in the cross
section. The new $\gamma n\rightarrow n\eta$ data obtained also at MAMI
with a $^3$He target \cite{Werthmuller:2013rba,Witthauer:2013} are consistent 
with the deuteron data apart from the absolute scale which is influenced 
by FSI. They were therefore not included into the fits. In addition to these 
data we use GRAAL data on the beam asymmetry $\Sigma$ for $\gamma n\to \eta n$ \cite{Fantini:2008zz}. 
The precision data from MAMI on $\gamma p\to\eta p$ \cite{McNicoll:2010qk} 
are discussed to clarify the underlying physical processes. The shallow dip at 1680\,MeV in the
$\gamma p\to p\eta$ cross section reported in \cite{McNicoll:2010qk} was described by introducing $N(1685)$ or by
assuming that there is a large $\omega$ coupling to the $J^P=1/2^-$
wave \cite{Anisovich:2013sva}. For pion production, the data on
$\gamma d\to \pi^0 np_{\rm spectator}$ and on $\pi^-p\to \gamma n$ listed in
Table 1 of \cite{Anisovich:2013jya} and the recent data from MAMI 
\cite{Dieterle:2014blj} on $\gamma d\to \pi^0 np_{\rm spectator}$ were used.

In all fits presented here, masses,  widths and coupling constants
for the decay of nucleon resonances are fixed except those for
helicity amplitudes of resonances and those for $t$ and $u$-channel
exchange amplitudes. The fixed values are taken from our fits to a
large body of $\pi N$ elastic scattering and pion and photo-induced
inelastic reactions (see \cite{Anisovich:2011fc,Anisovich:2013vpa}
for references to the data included).

\section{\label{sim}Simulations}
Before we present the results of the  partial wave analysis, we
present some simulations which demonstrate what to expect given a
particular hypothesis. The simulations are close to the experimental
observations but the underlying model is simpler and focuses on
specific aspects of the reaction.

\subsection{\label{simsp}\boldmath Simulation of the narrow structure with $J^P=1/2^-$ and $J^P=1/2^+$ states}

\begin{figure}[thb]
\centerline{\epsfig{file=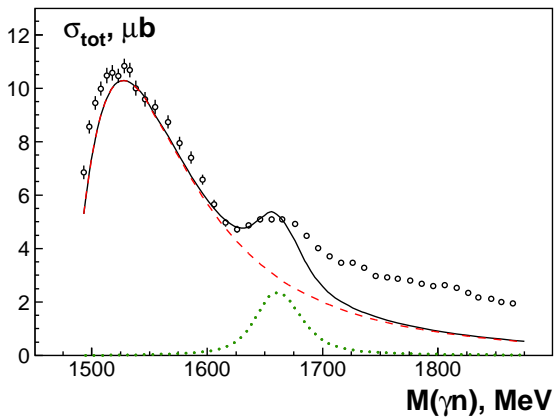,width=0.42\textwidth}}
\caption{\label{tot_sim} (color online)  Simulation of the total
cross section (solid line) with contributions from  $J^P=1/2^-$ (dashed
curve, red) and $J^P=1/2^+$ (dotted curve, green).}
\end{figure}

First we study the interference between a dominant  $J^P=1/2^-$ wave
and a narrow resonance in the $J^P=1/2^+$ wave. A typical example is
shown in Fig.~\ref{tot_sim}. Two contributions are present:
$N(1535)$ provides a strong $J^P=1/2^-$ wave at low energies, a narrow
$J^P=1/2^+$ resonance with a mass of 1680\,MeV forms the second peak in
the total cross section.

\begin{figure}[ht]
\centerline{\epsfig{file=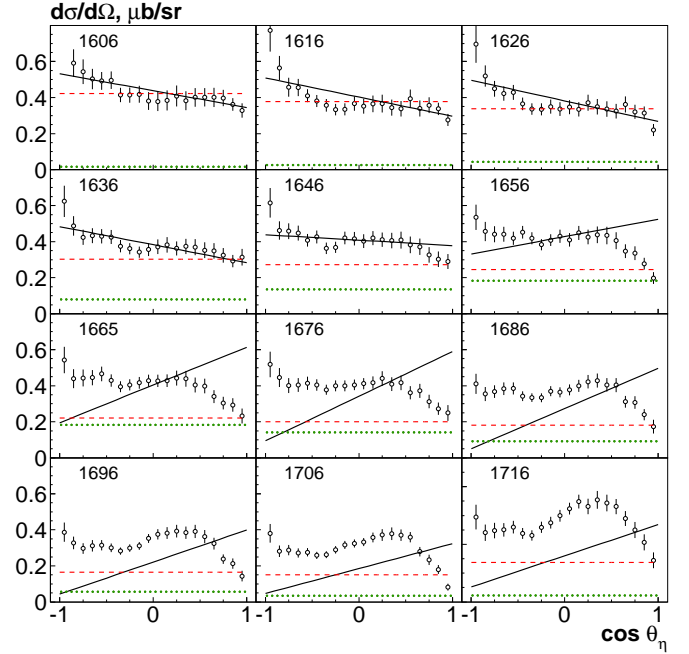,width=0.48\textwidth}}
\caption{\label{dcs_sim} (color online) Differential cross section
for the eta photoproduction with (solid curve) and  without
interference between $J^P=1/2^-$ (dashed curve, red) and $J^P=1/2^+$
(dotted curve, green) waves. The numbers correspond to the central mass of a bin (in MeV).}
\end{figure}

\begin{figure}[ht]
\centerline{\epsfig{file=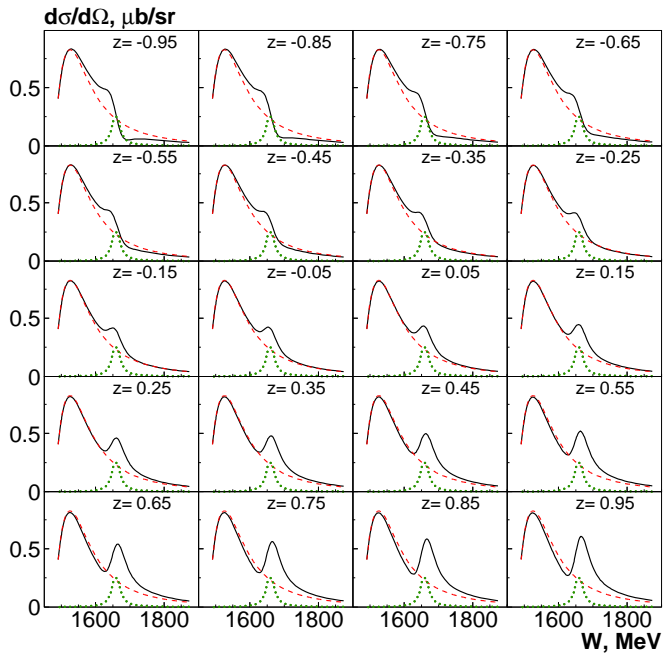,width=0.48\textwidth}}
\caption{\label{en_sim} (color online) Energy distributions at fixed angles (in bins of $z=\cos\Theta_\eta$) in the case of the interference between $J^P=1/2^-$ and $J^P=1/2^+$ states. The
contributions of the $J^P=1/2^-$ partial wave are shown with dashed (red) curves
and $J^P=1/2^+$ with dotted (green) curves.}
\end{figure}

Without  interference, both partial waves produce uniform angular
distributions, see the dashed (red) and dotted (green) curves in
Fig.~\ref{dcs_sim}. The interference between the two waves generates
a linear angular distribution; its slope depends on the phase
between the $J^P=1/2^-$ and $J^P=1/2^+$ waves. Overall, the area below the
solid line is larger than the area below the dashed line: the narrow
$J^P=1/2^+$ resonance brings in additional intensity. However, in most
energy bins there are angular regions where the solid line is below 
the dashed one. These are regions of strong destructive interference.

The excitation  functions at fixed angles are shown in
Fig~\ref{en_sim}. The distributions show either a peak at the mass
of the $J^P=1/2^+$ resonance or a diffractive pattern. We will see that
the data are inconsistent with these assumptions. Note that the total 
cross section can be calculated from the differential cross section
and from the excitation function by calculating the mean of all $\cos\theta$ bins  
and multiplying with $4\pi$.

\subsection{\label{simss}Simulation of the interference pattern}

\begin{figure}[htb]
\centerline{\epsfig{file=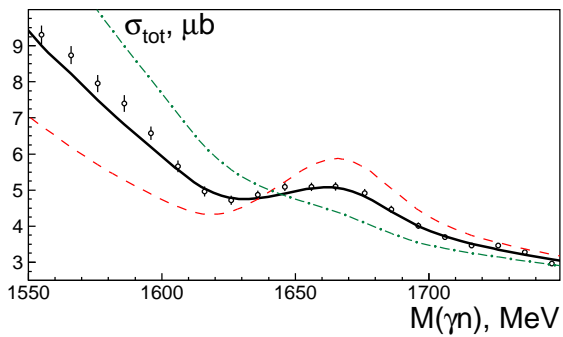,width=0.42\textwidth}}
\caption{\label{hel_sim} (color online) The  total cross section for
$\gamma n\to \eta n$ \cite{Werthmuller:2013rba} with our best fit
(solid curve) and with predictions in which the helicity amplitude
$A_{1/2}^n$ of $N(1650)1/2^-$ is multiplied by a factor 2 (dashed,
red) or -1 (dashed-dotted, green), respectively.}\vspace{-2mm}
\end{figure}

Figure~\ref{hel_sim} shows a simulation of the  effect of the
interference between the two $J^P=1/2^-$ resonances. The solid curve
represents our best fit (as discussed below). The $J^P=1/2^-$ wave is
represented by a K-matrix with three poles and smooth background \cite{Anisovich:2008wd} (which contributes less than 10\% to the cross section). Then we have multiplied the
$A_{1/2}^n$ helicity amplitude of $N(1650)1/2^-$ with a factor 2 or
(-1). This has a very significant impact on the predicted cross
section. When the $N(1650)1/2^-$ helicity amplitude is increased
(from $0.019$ to $0.038$\,$10^{-3}$\,GeV$^{-1/2}$), the peak
structure grows significantly while only a small effect remains
visible when the sign of the helicity amplitude is changed.

%
%
%
%

\subsection{\label{simcusp}Simulation of a cusp}

\begin{figure}[thb]
\centerline{\epsfig{file=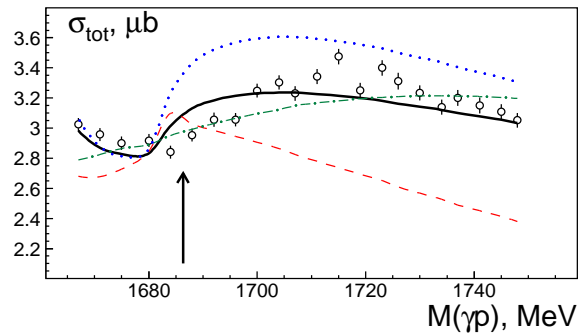,width=0.42\textwidth}}
\caption{\label{en_cusp}(Color online) Total  cross section for
$\gamma p\to \eta p$. Solid (black) curve: best fit; dashed-dotted
(green) curve: fit with zero coupling to the $K\Sigma$ final state;
dotted (blue) curve: coupling of $N(1650)1/2^-\to K\Sigma$ doubled;
dashed (red) curve: coupling of $N(1650)1/2^-\to K\Sigma$ with negative
sign. The arrow indicates the position of the $K\Sigma$ threshold.}
\end{figure}

We show the effect of the opening of a new  channel using data on
$\gamma p\to \eta p$. For this reaction a small dip in the total
cross section at about 1.68\,GeV was reported
\cite{McNicoll:2010qk}. The structure was soon assigned to the
narrow nucleon resonance $N(1685)$ \cite{Kuznetsov:2011pe}. A
detailed study showed that the structure can be described well when
the $N(1685)$ nucleon resonance with spin-parity $J^P=1/2^+$ is
added to the list of resonances used in the BnGa partial wave
analysis \cite{Anisovich:2013sva}. However, an equally good fit was
obtained when taking into account the opening of the reaction
$\gamma p\to p\omega$ at 1720\,MeV.

At the time when the study \cite{Anisovich:2013sva} was made, data on
photoproduction of $\omega$ mesons were not yet included in the BnGa
analysis. With such data included, it turned out that the $p\omega$
coupling of the $J^P=1/2^-$ partial wave would need to be considerably
larger than the data on $\gamma p\to p\omega$ suggest. Hence we decided
to study the effect of the opening of the reaction $\gamma p\to K\Sigma$.
In a recent analysis, we reported an ambiguity in the signs of $N\to
K\Sigma$ coupling constants \cite{Anisovich:2013vpa}. The new
solution (BnGa2013-02) had a much more significant $J^P=1/2^-$
contribution than the solution (BnGa2011-02M), see Fig.~12 of
\cite{Anisovich:2013vpa} and is in an excellent agreement with the
solution found by R\"onchen {\it et al.} \cite{Ronchen:2012eg}. Based on
the new solution, the total cross section for $\gamma p\to \eta p$
should exhibit a clear structure at the $K\Sigma$ threshold which
is shown in Fig.~\ref{en_cusp}. This is the case, indeed. The figure
shows the expected distribution (solid curve) and curves which are
predicted for the case of vanishing $K\Sigma$ coupling, for a coupling
multiplied by a factor 2, and for a factor {\nobreak{(-1)}}.
The solid line does not match the data points exactly but the effect of
the $K\Sigma$ threshold is clearly seen. Note that the errors correspond
to statistical errors only. The importance of the $K\Sigma$ coupling for
the description of $J^P=1/2^-$ structure around 1700 MeV wave was stressed
by M.~D\"oring (see \cite{Doring:2012zz} and references therein).

\section{\label{pwa}Partial wave analysis of the data}

We start our analysis from the solutions  reported in
\cite{Anisovich:2013jya}. These solutions were obtained by fitting
almost the full data base on $\gamma n\to \pi N$, $\pi^-p\to \gamma n$,
and on $\gamma n\to \eta n$.

\begin{figure}[thb]
\centerline{\epsfig{file=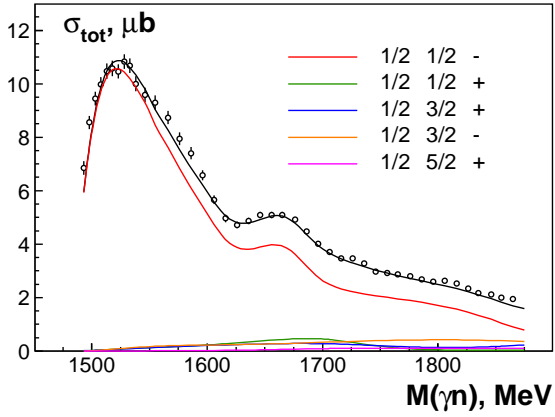,width=0.42\textwidth}}
\caption{\label{tot_s11}(color online) The  total cross section for $\gamma n\to
\eta n$ and the contributions from partial wave with isospin $1/2$ and different spin-parities.  
There is no narrow $N(1685)$
admitted in the fit. The structure at this mass is described by
interference within the $J^P=1/2^-$ wave.}
\end{figure}

\begin{figure}
\centerline{\epsfig{file=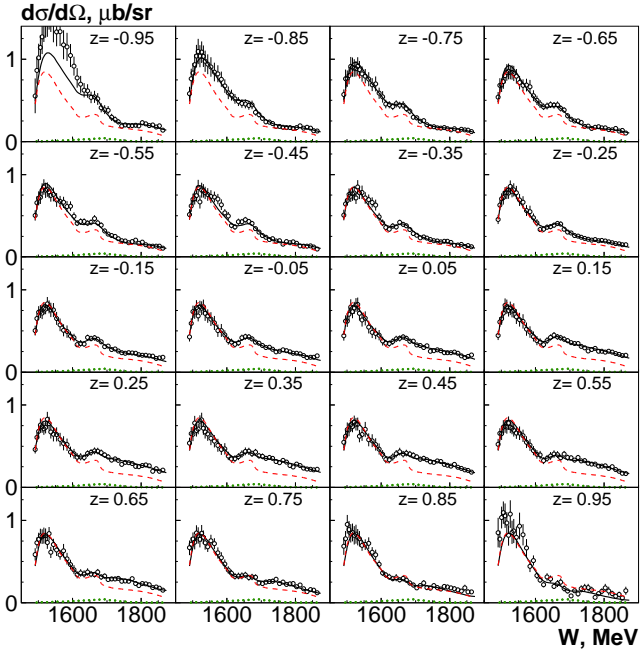,width=0.48\textwidth}}
\caption{\label{dcs_energy} (color online) Excitation
function for $\gamma n\to \eta n$ at fixed angles (in bins of $z=\cos\Theta_\eta$). Only statistic
errors are  shown. There is no narrow $N(1685)$ admitted in the
fit. The fit is represented by the full (black) curve; the $J^P=1/2^-$ wave by the dashed (red) curve. 
The contributions from partial wave with isospin $1/2$ and different spin-parities are shown by colored curves.}
\end{figure}

\subsection{Fits with no narrow nucleon resonance}

\begin{figure}[pt]
\centering
\epsfig{file=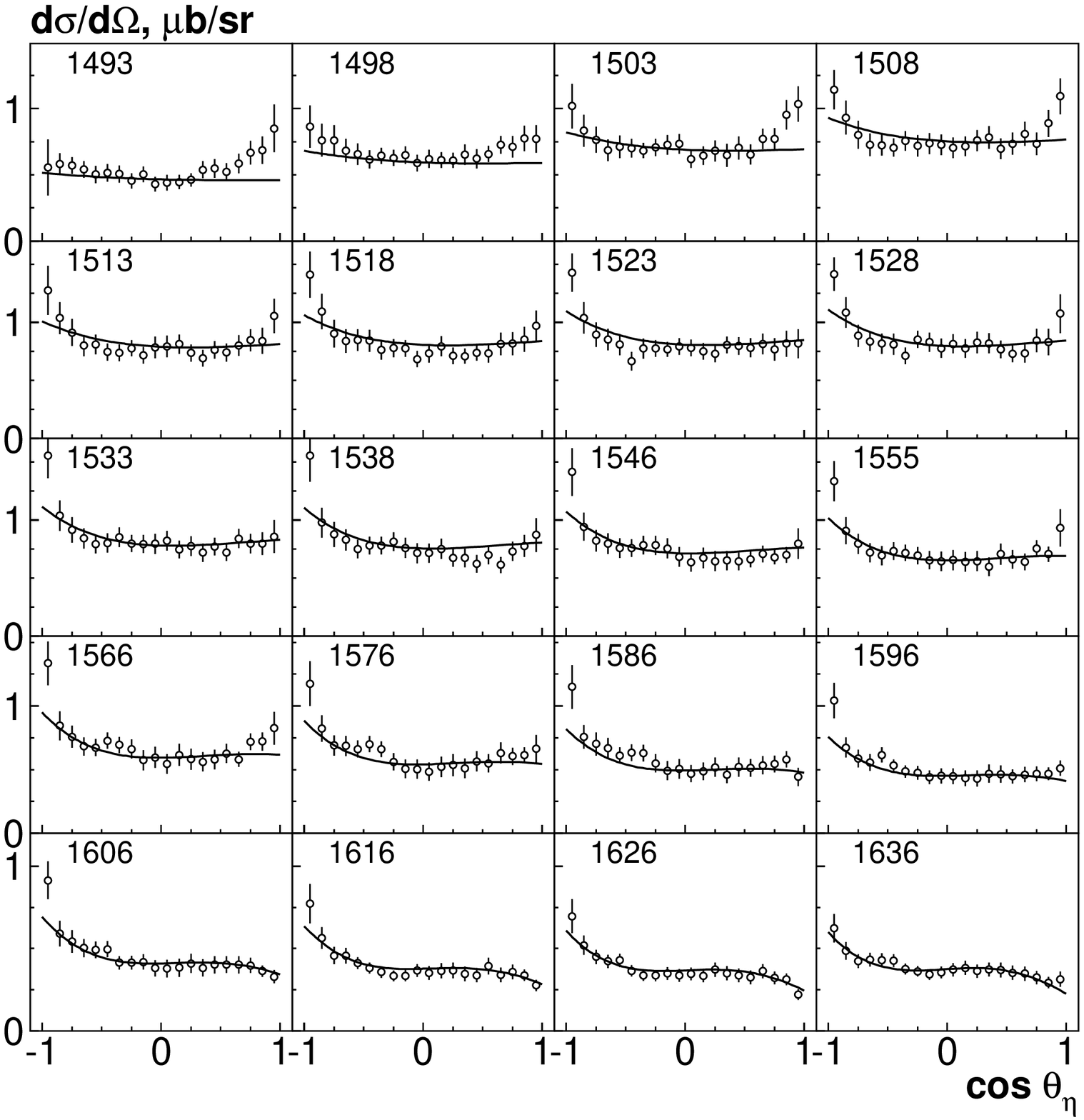,width=0.48\textwidth}\\[1ex]
\epsfig{file=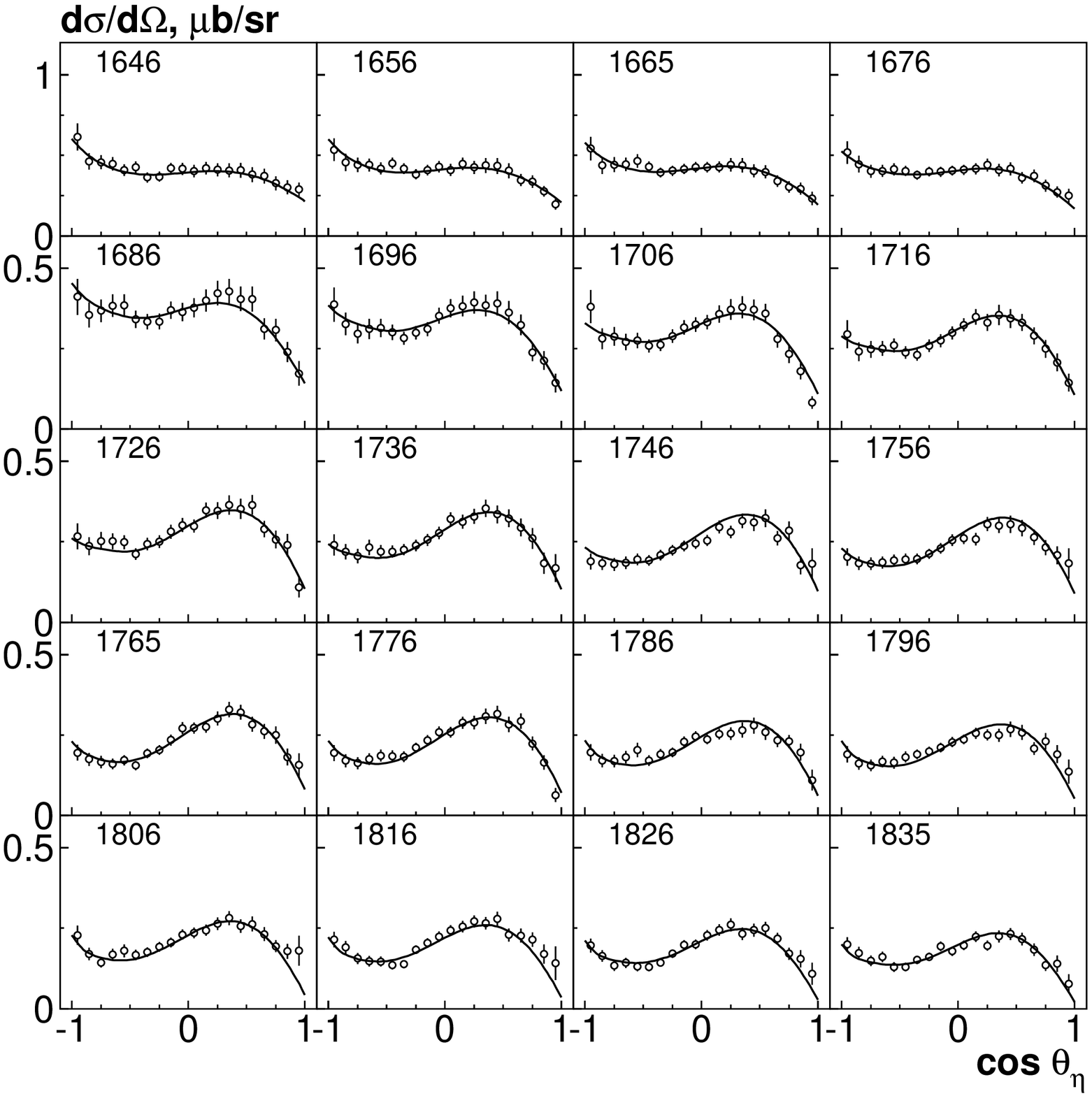,width=0.48\textwidth}
 \caption{\label{dcs_free_1} (color online) Differential
cross sections for $\gamma n\to \eta n$. Only statistic errors are
shown. There is no narrow $N(1685)$ admitted in the fit.\vspace{2mm}}
\end{figure}

First, we fitted the data with  conventional nucleon resonances
only. Figure~\ref{tot_s11} shows the total cross section with the fit
and the most significant partial wave contributions. Clearly, the
$J^P=1/2^-$ wave is dominant; the fit finds small contributions from
the $J^P=1/2^+$,  $3/2^-$, $3/2^+$, and $5/2^-$  waves. In the fit,
we use statistical and systematic errors added quadratically.
The fit returns $\chi^2$ values per data point which are often
smaller than $1$. This is not surprising since several sources of
systematic uncertainty vary slowly with energy or are constant.
Using the statistical errors only, typical $\chi^2$ values per data
point are slightly above 3, indicating the need for error contributions
beyond the statistical errors. The conclusions of the paper are not
affected when the systematic errors are included or neglected.

\begin{figure*}[pt]
\begin{center}   
\begin{tabular}{cc}
\epsfig{file=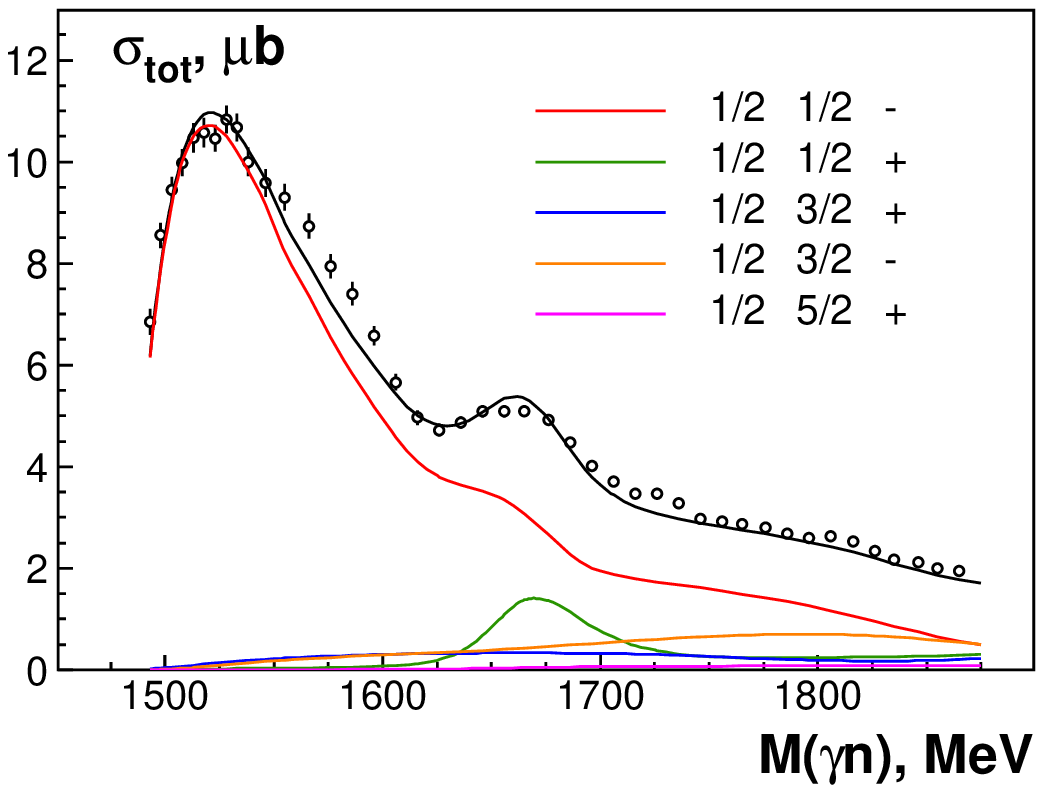,width=0.44\textwidth}&
\epsfig{file=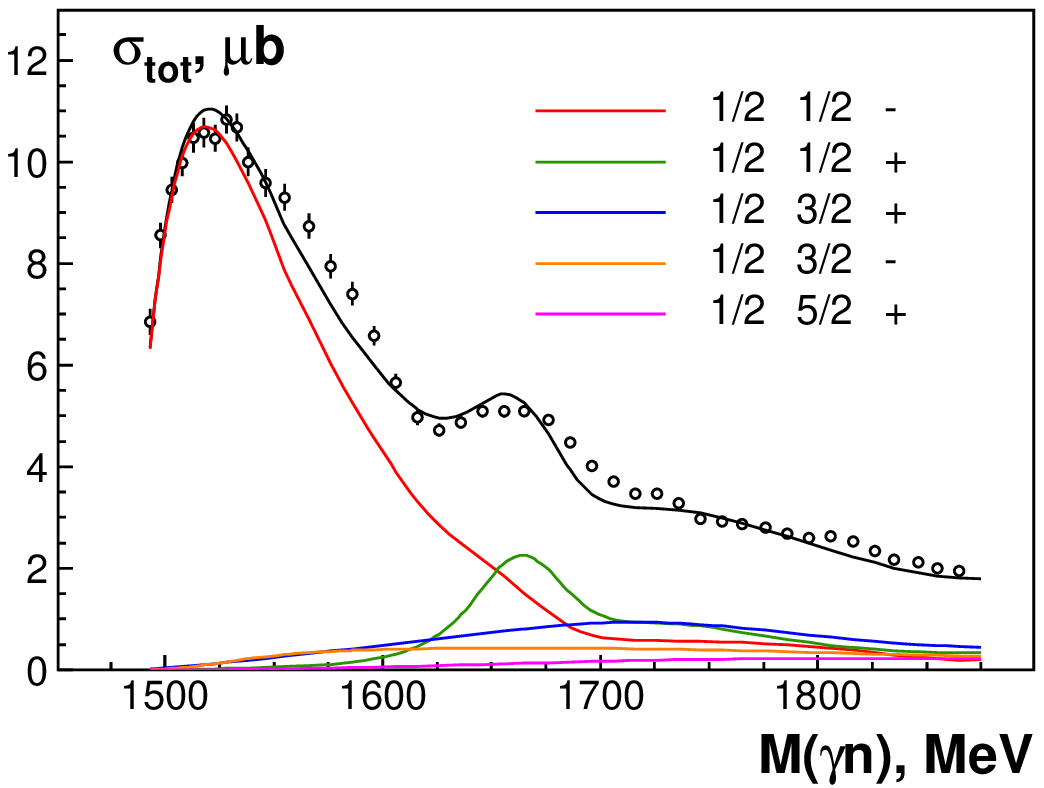,width=0.44\textwidth}\\
\epsfig{file=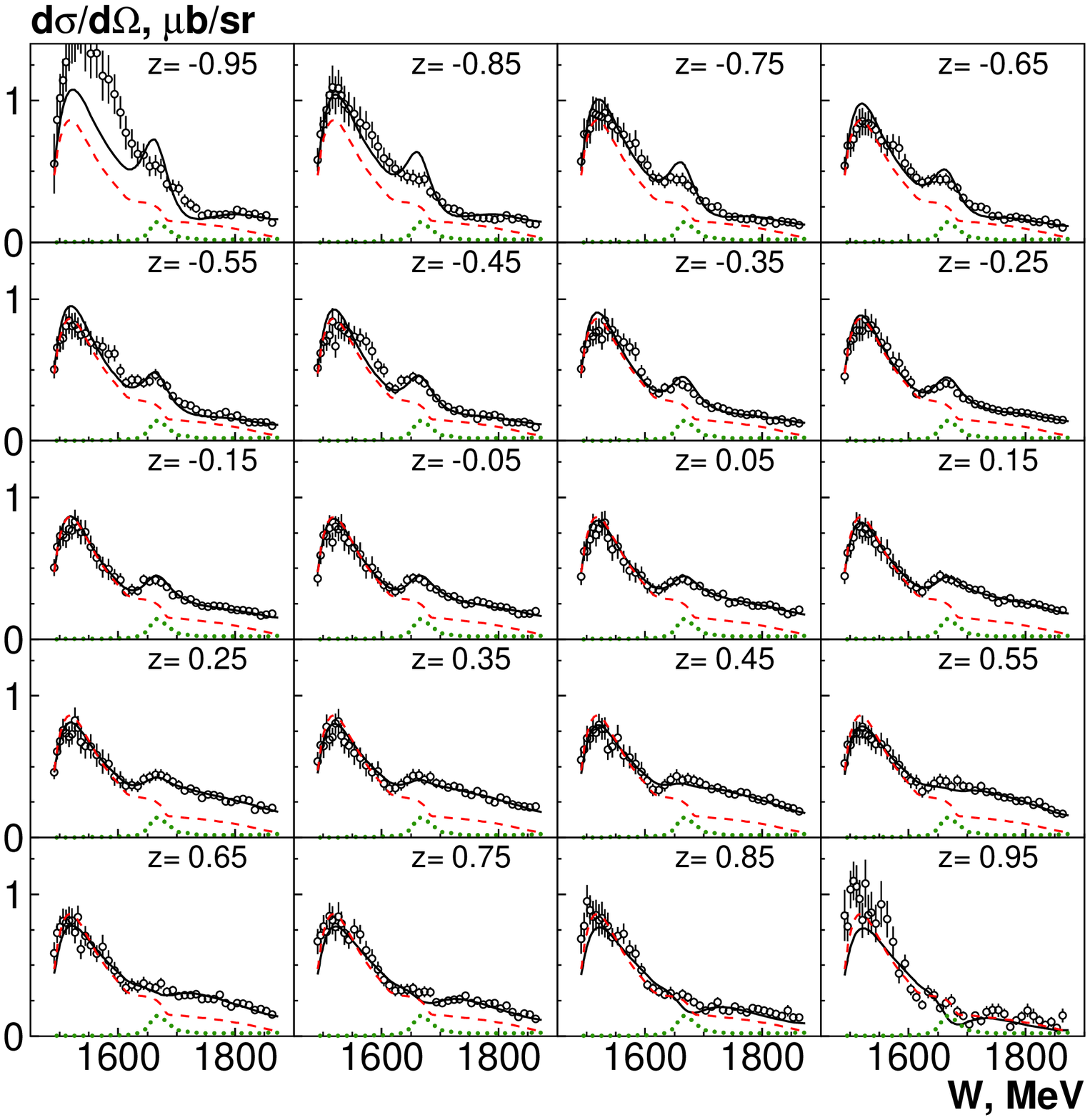,width=0.48\textwidth}&
 \epsfig{file=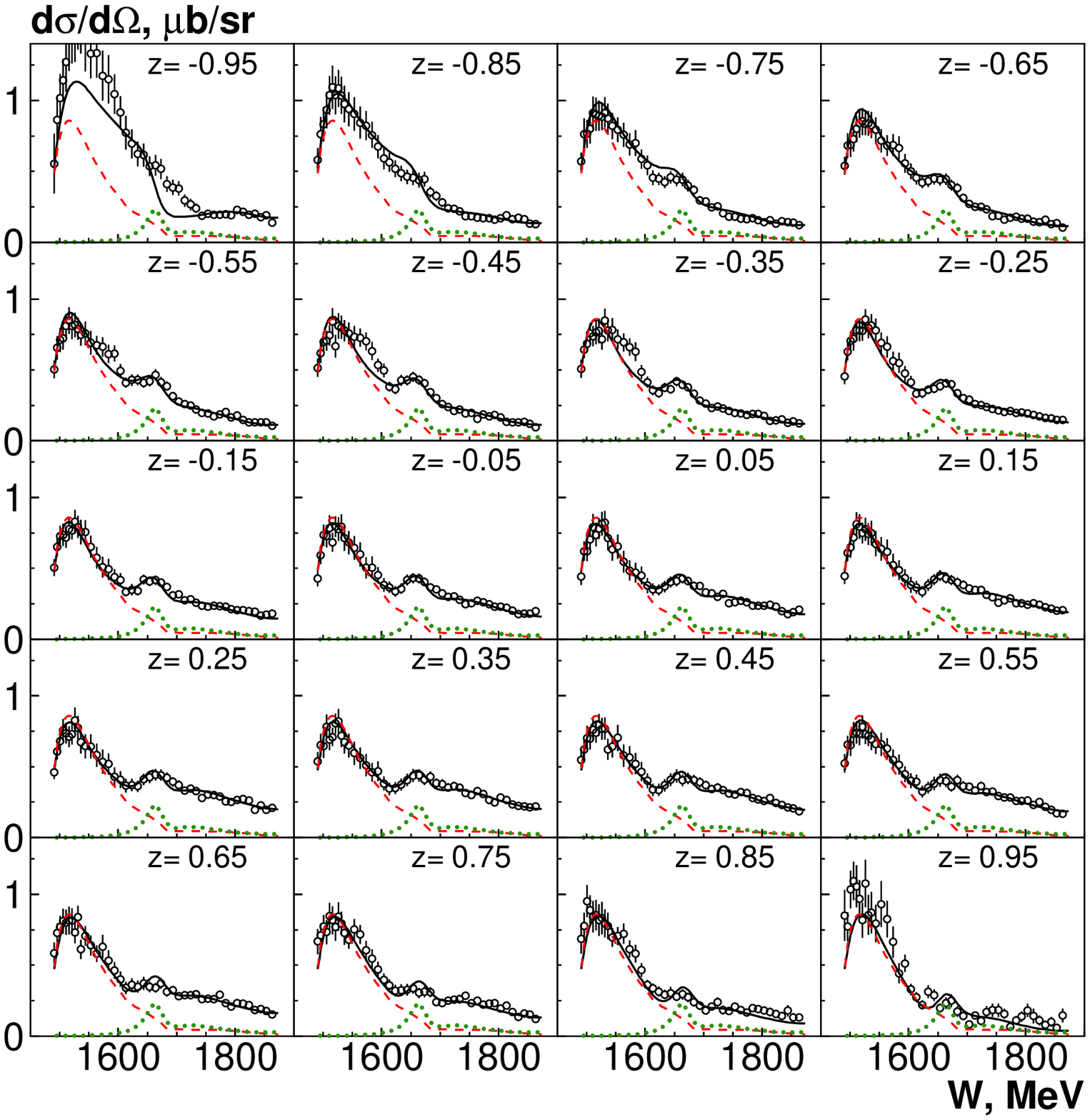,width=0.48\textwidth}
\end{tabular}
\vspace{-2mm}
\end{center}
\caption{\label{tot_p11} (color online)Total cross section (top) 
and excitation functions at fixed angles (bottom). Only statistical errors are shown. 
In the fit a narrow narrow $J^P=1/2^+$ resonance at 1685\,MeV is imposed.  
Left: $\tilde a>0$; right: $\tilde a<0$ (see text). }
\end{figure*}

The quality of the fit can be judged by inspecting Figs.~\ref{dcs_energy} 
and \ref{dcs_free_1}. A $\chi^2$ of 0.91 per
data point was achieved. A large fraction of the $\chi^2$ stems from
the most backward $\eta$ production angle ($z=\cos\theta=-0.95$). In
Fig.~\ref{dcs_energy}, a significant excess of data compared to the
fit is seen in this angular range at low energies ($1520<W<1620$\,MeV).

The angular distributions (Fig.~\ref{dcs_free_1}) suggest that this
excess might be artificial. In this mass range, most backward data points seem
anomalously high. The detection efficiency for these points (see Fig.~13 in
\cite{Werthmuller:2014}) is much lower than for the second point in the angular
distributions and the systematic uncertainty is larger.
This small excess also explains the small discrepancy between data and fit in the total cross section around 1580\,MeV (Figs.~\ref{tot}, \ref{hel_sim}, and \ref{tot_s11}). 
We conclude that an excellent fit to the data \cite{Werthmuller:2014,Werthmuller:2013rba} can be achieved without introducing a narrow resonance $N(1685)$. In the 1610 - 1710\,MeV mass region, the $\chi^2$ per data point is 0.48.

\subsection{\boldmath Fits imposing a narrow nucleon resonance $N(1685)$}

In the next step we investigated the scenario with a narrow $J^P=1/2^+$ 
resonance interfering with the $J^P=1/2^-$ partial wave.
We added a $J^P=1/2^+$ resonance in the 1680 MeV\,mass region. A fit with 
free width and real $\eta n$ coupling converged to a solution with a very broad 
resonance (more than 200 MeV) and a very weak coupling. The improvement in 
$\chi^2$ was negligible.

We then imposed contributions from a $N(1685)$ resonance with properties 
corresponding to the phenomenologic fits in \cite{Werthmuller:2013rba}: 
mass $M=1670\pm 5$ MeV, width $\Gamma= 30\pm 15$ MeV, and $\sqrt{Br(\eta
n)}A_n^{1/2}=\tilde a$ [GeV$^{-\frac12}$\,10$^{-3}$] = $(12.3\pm
0.8)$ [GeV$^{-\frac12}$\,10$^{-3}$]. For $\tilde a$  we assumed,
alternatively, to have a positive or a negative sign. (When complex
values were admitted, the overall $\chi^2$ improved slightly but the
$\chi^2$ restricted to the 1620 - 1720\,MeV region was worse.)

\begin{table*}[t]
\caption{\label{hel_coup} (color online) Solutions  with and without a contribution
from a narrow resonance in the $J^P=1/2^+$wave. The product of helicity
coupling and $\eta n$ branching ratio is given in units of
GeV$^{-\frac 12}10^{-3}$. The $\chi^2_{dcs}$ for the differential
cross section is calculated in the region $1.610-1710$\,MeV. The
$\chi^2_{\Sigma}$ for the beam asymmetry $\Sigma$ is given
separately. Results are given for two fits using only the
statistical errors or the total errors.\vspace{2mm}}
\begin{center}
\renewcommand{\arraystretch}{1.4}
\begin{tabular}{lcccccccc}
\hline\hline
Fit & Mass & Width &$\sqrt{Br(\eta n)}|A_n^{1/2}|$ & Phase
&$\chi_{dcs}^2/200$ & $\chi_\Sigma^2/80$ &$\chi_{dcs}^2/200$ & $\chi_\Sigma^2/80$ \\
& & & & &\multicolumn{2}{c}{stat. + syst. errors} &\multicolumn{2}{c}{stat. errors only}\\
\hline
$J^P=1/2^-$& -   & -  & -  & -           &0.48 & 1.81 & 3.10 &2.10 \\
$J^P=1/2^+$& 1671& 35 &-12 & $0^o$       &1.34 & 2.80 & 9.35 & 2.92 \\
$J^P=1/2^+$& 1669& 35 &+12 & $0^o$       &1.47 & 2.71 & 7.66 & 3.02\\
$J^P=1/2^+$& 1671& 35 &-12 & $20^o$      &1.50 & 2.50 & 9.33 & 2.90\\
$J^P=1/2^+$& 1674& 35 & 5 & $0^o$        &0.55 & 1.98 & 3.40 & 2.50\\
$J^P=1/2^+$& 1671& 35 & -3 & $0^o$       &0.54& 1.95 & 3.30 &2.55 \\
\hline\hline
\end{tabular}
\renewcommand{\arraystretch}{1.0}
\end{center}\vspace{-2mm}
\end{table*}
The total cross sections and the excitation functions in different bins of the $\eta$ production
angle for the solutions with positive and negative product couplings are shown in 
Fig.~\ref{tot_p11}.

The difference  between the two solutions is seen very well. The
solution with $\tilde a>0$ shows a strong peak at 1685\,MeV for
backward $\eta$ mesons, while the solution with $\tilde a<0$
exhibits a diffractive structure. For forward $\eta$, the opposite
holds true. The narrow $J^P=1/2^+$ resonance produces an asymmetry
which is not supported in the data. For $\tilde a<0$ this asymmetry
is partly interpreted in the fit by an increase in the contribution
from the $P_{13}$ partial wave. The narrow $N(1685)$ interferes with
the broad $N(1710)$, destructively in Fig.~\ref{tot_p11}, left panel, and
constructively in Fig.~\ref{tot_p11}, right panel. Interference with other
waves cannot be observed after integration over the full angular
distribution.

The data clearly disfavor the scenario with a narrow $J^P=1/2^+$
resonance. The resulting interference with the $J^P=1/2^-$ wave 
produces the expected forward - backward asymmetry in the angular
distributions, which is not reflected in the experimental data.
In a further step we have determined upper limits for the quantity $\tilde a$.
We find that the description of the differential cross section is
still compatible with the data for $-3<\tilde a< +5$.

\subsection{Comparison of the quality of the two fits}

In Table~\ref{hel_coup} we compare the quality of  various fits. The
best fit is achieved when no narrow $N(1685)$ is imposed. In the
table we give the $\chi^2$ per data point for the differential cross
section~\cite{Werthmuller:2013rba} and for the beam
asymmetry~\cite{Fantini:2008zz}. The mass range for which the
$\chi^2$ is calculated is restricted to $1610-1710$\,MeV, the range
which is most relevant for the existence of $N(1685)$. Fits are
performed using the statistical errors only and the statistical and
systematic error added quadratically.

\begin{figure*}[t]
\begin{center}
\epsfig{file=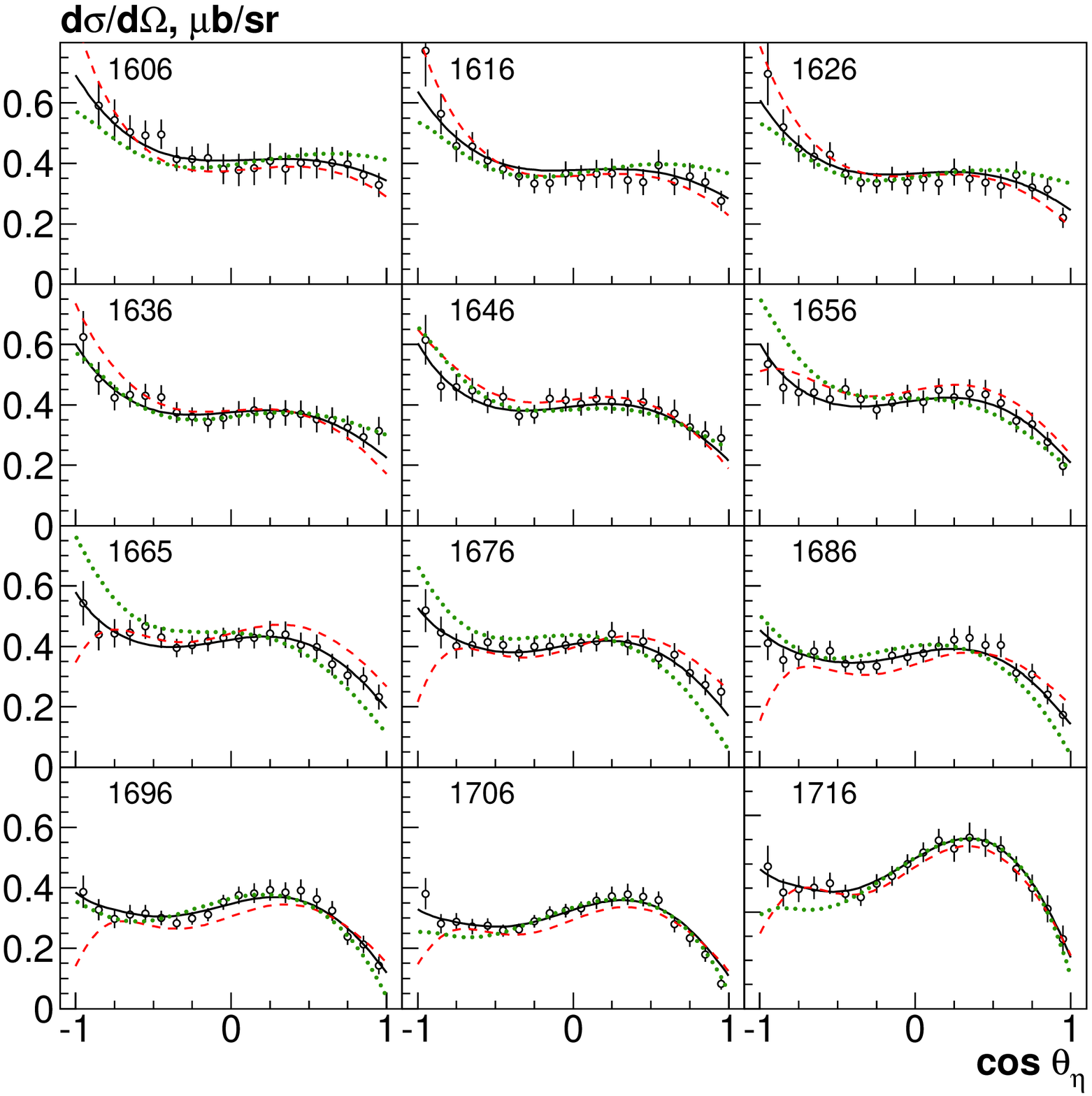,width=0.49\textwidth}
\epsfig{file=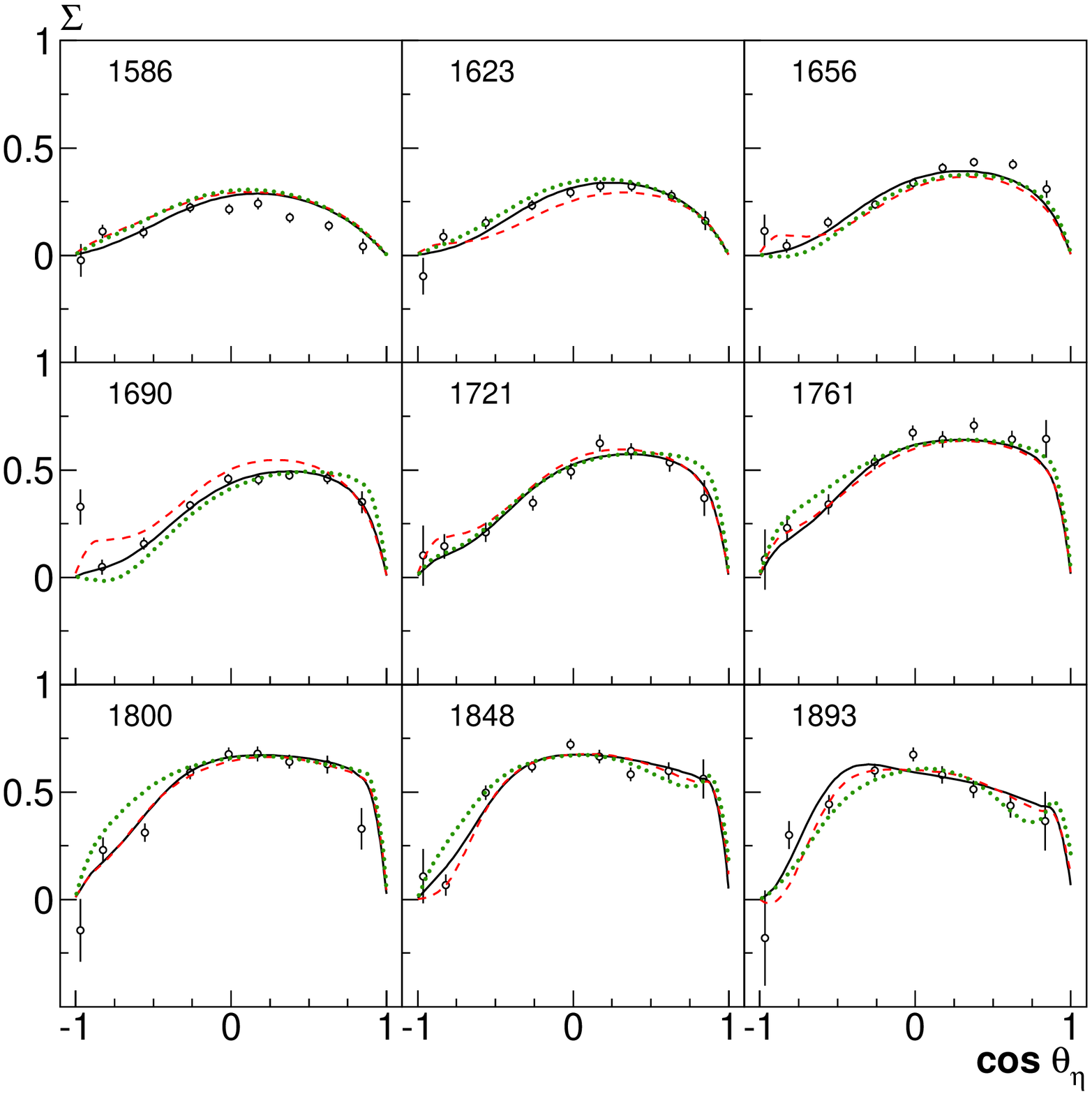,width=0.49\textwidth}
\vspace{-2mm}\end{center} \caption{\label{dcs_sel} (color online) Differential
cross section (left) and beam asymmetry (right) for the eta
photoproduction off neutrons. The numbers correspond to the central mass of a bin (in MeV). The solution with $J^P=1/2^-$ interference is shown as black curves, the solution with
$N(1680)1/2^+$ state with negative $\eta n$ coupling as
dashed (red) curves, and the solution with positive coupling as dotted (green) curves.\vspace{2mm}}
\end{figure*}

The fit in which no narrow $N(1685)$ is imposed  gives the best
$\chi^2$ and is our favored fit. It is shown in Figs.~\ref{dcs_sel}
as solid curve. The differential cross sections are perfectly
described; the beam asymmetry at 1586\,MeV shows a few points which
are missed by the fit but in the next energy bin, fit and data are
already fully consistent. In contrast to these findings, the fit
with enforced contributions from a narrow $N(1685)$ exhibits
significant deviations from data. In these fits, the $N(1685)$ mass
is constrained by $1660 < M < 1710$\,MeV, the width by $\Gamma
<35$\,MeV while the product branching ratio is fixed to $\tilde
a=\sqrt{Br(\eta n)}A_n^{1/2}=\pm 12$\,GeV$^{-\frac12}$\,10$^{-3}$.
Both solutions (with positive and negative $\tilde a$)
provide a strong backward-forward asymmetry in the angular
distributions which is not supported by the data. To compensate this
asymmetry the fit increased the contribution from the $J^P=3/2^+$
partial wave. However, it leads to a deterioration for the
description of the GRAAL beam asymmetry data. The differential cross
section in the mass region 1600 - 1720 MeV and the GRAAL beam asymmetry
data are compared with different fits in Fig.~\ref{dcs_sel}. The fits
return the masses listed in Table~\ref{hel_coup} while the width goes to
the boundary value. In one of the fits (not shown), $\tilde a$ was
defined as a complex number with $|\tilde a=12|$\,GeV$^{-\frac12}$\,10$^{-3}$.
The fit gave a mar\-ginal overall improvement.

Finally, we determine upper limits for contributions. For $\tilde
a=5$, first visible deviations between data and fit show up and the
total $\chi^2$ increases by slightly more than 25, corresponding to
$5\sigma$. Similar discrepancies are observed for $\tilde a=-3$. We
conclude that -- if a narrow $N(1685)$ exists in the $J^P=1/2^+$ wave
-- its production and decay branching fractions must obey $-3<\tilde
a<5$.

Numerically, the fit converges to helicity  amplitudes given in
Table~\ref{helamp}. Obviously, the two $J^P=1/2^-$ resonances interfere
differently. The helicity ratios for the $\eta$-photoproduction off protons, recent data on the double-polarization variables $F$ and $T$ from MAMI~\cite{Akondi:2014ttg}
and on $E, G, T, P, H$ from ELSA~\cite{Muller:2015tbd} are included.  

\begin{table*}[pt]
\caption{\label{helamp}Helicity amplitudes determined from a fit without a narrow $N(1685)$ resonance. 
The T-matrix couplings are those which are listed in the RPP. K-matrix couplings are reproduced as small numbers.
The results of a fit to the 2002 data using the Single-Quark-Transition model \cite{Burkert:2002zz} are shown for comparison.}
\begin{center}            
\renewcommand{\arraystretch}{1.5}
\begin{tabular}{|l|cccl|l|cccl|}
\hline\hline
&& $N(1535)1/2^-$ & $N(1650)1/2^-$ && && $N(1535)1/2^-$ & $N(1650)1/2^-$ &\\\hline
  \multirow{3}{*}{\boldmath$p$}&T-matrix&   $0.114\pm 0.008$ &   $0.032\pm 0.007$ &GeV$^{-1/2}$ &  \multirow{3}{*}{\boldmath$n$} &T-matrix&   -$0.095\pm 0.006$ &   $0.019\pm 0.006$ &GeV$^{-1/2}$\\[-1ex]
 &Phase&   10\er5$^\circ$ & -2\er11$^\circ$  & &  &Phase&   8\er5$^\circ$ &  0\er15$^\circ$  &\\[-1ex]
&K-matrix&\scriptsize   $0.096\pm 0.007$           &\scriptsize  $0.075\pm 0.007$  & & &K-matrix&\scriptsize   -$0.120\pm 0.006$           &\scriptsize  -$0.052\pm 0.006$   &\\[-1.2ex]    
 &SQT&\scriptsize   $0.097\pm 0.007$           &\scriptsize  $0.053\pm 0.004$         &   &&SQT&\scriptsize   -$0.090\pm 0.006$           &\scriptsize  -$0.031\pm 0.003$         &\\   
\hline\hline
\end{tabular}
\end{center}\vspace{2mm}
\end{table*}

For the proton the helicity couplings of both states have like sign, for the neutron opposite sign. However, the hadronic phase involved in the $N^{\star}\rightarrow N\eta$ decay of the two 
states (with respect to pion production) is predicted to be positive for the $N(1535)1/2^-$ 
and negative for the $N(1650)1/2^-$~\cite{Capstick:1993kb}; experimentally, the $\pi N\to N^*\to \eta N$ transition residue has a phase of $-(76\pm5)^\circ$ for $N(1535)1/2^-$ and $+134\pm10^\circ$ for $N(1650)1/2^-$ \cite{Anisovich:2011fc}; their relative phase is hence $+(210\pm12)^\circ$ close to the predicted $180^\circ$. Hence the resulting interference is destructive for the reaction $\gamma p\to \eta p$ and constructive for $\gamma n\to \eta n$. 
It was pointed out in \cite{Boika:2014aha} that the $N(1650)1/2^-$ helicity amplitude is at variance with model predictions (see, e.g., \cite{Anisovich:2013jya} for references to predictions), and in particular that the positive sign of the $N(1650)1/2^-$ helicity amplitude is unexpected. It was shown that the value implies that the $N(1650)1/2^-$ must have a large $s\bar s$ component. 

This statement holds for the T-matrix coupling constants which give the helicity amplitudes at the pole position of the ``dressed'' resonance. These are complex numbers. In Table~\ref{helamp} we also list the K-matrix helicity amplitudes. The K-matrix pole characterizes the position of the pole when all decay modes are switched off. Thus, the K-matrix poles can be interpreted as pole position of the ``undressed'' resonance. In the neighborhood of important thresholds - in this case of the $K\Lambda$ threshold - T-matrix and K-matrix coupling constants can differ substantially. The helicity amplitudes of the ``undressed'' $N(1650)1/2^-$resonance are real numbers. The ``undressed'' helicity amplitudes are in good qualitative agreement with the predictions of the Single-Quark-Transition model~\cite{Burkert:2002zz} and do not necessitate a large $s\bar s$ component in the  $N(1650)$ $1/2^-$ wave function.

\subsection{\label{pineutron}\boldmath Interference in the $J^P=1/2^-$ wave in $\gamma n\to \pi^0 n$}

\begin{figure*}[t]
\centerline{\epsfig{file=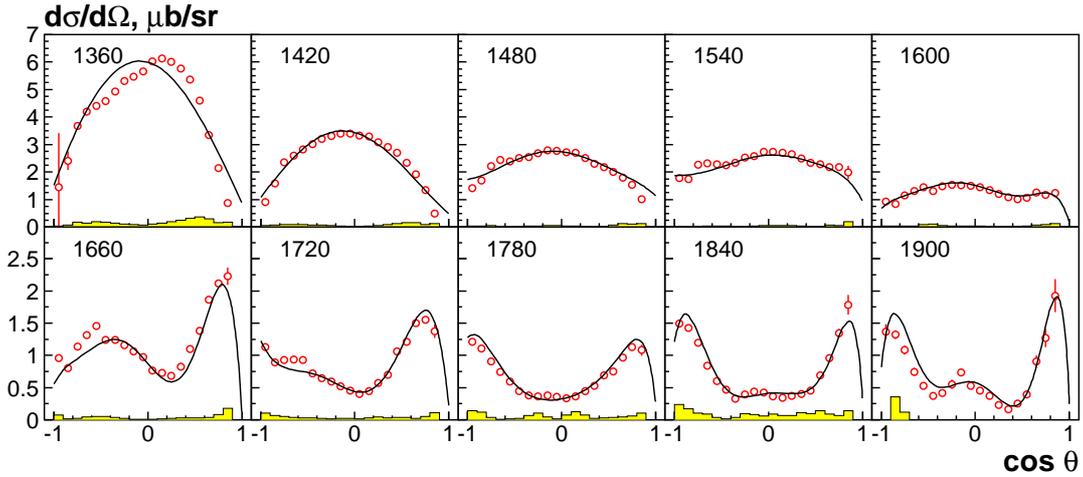,width=0.80\textwidth}}
\caption{\label{npi0diff}Differential cross sections for
$\gamma n\to \pi^0 n$~\cite{Dieterle:2014blj}. The solid curve shows the BnGa fit
to the data. The statistical errors are mostly smaller than circles representing the data.
The systematic errors are shown as yellow histograms.  }
\end{figure*}

\begin{figure}[t]
\vspace{3mm}
\centerline{\epsfig{file=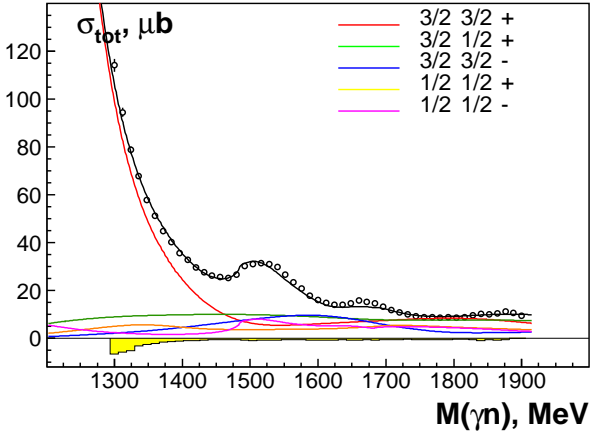,width=0.44\textwidth}}
\caption{\label{npi0} (color online) The  total cross section for
$\gamma n\to \pi^0 n$~\cite{Dieterle:2014blj}. The black curve shows the BnGa fit
to these data, the colored curves the contributions from different partial waves $J^P$ with
isospin $1/2$ and $3/2$.}
\end{figure}

We now ask if a trace of the controversially discussed $N(1685)$ can be found in the reaction $\gamma n\to \pi^0 n$. 
In this case, the final state consists of a proton, a neutron, and a pion (instead of an $\eta$). The data are much stronger influenced by final state interactions than $\eta$-photoproduction \cite{Dieterle:2014blj}. With this warning we show in Fig.~\ref{npi0diff} the differential  cross section and in Fig.~\ref{npi0} the total cross section  for $\gamma n\to \pi^0 n$ measured with the Crystal Ball/TAPS calorimeter at the MAMI accelerator~\cite{Dieterle:2014blj}. At low energy, the reaction is dominated by $\Delta(1232)$ 
production; its tail exceeds all other contributions up to an invariant mass of 1450\,MeV. The peak in the second resonance region is mainly assigned to $N(1520)3/2^-$, the small peak below 1700\,MeV to $N(1680)5/2^+$. The figure also displays the contribution of the $J^P=1/2^-$ wave to the total cross section. The contribution is small and shows the opening of the $\eta n$ threshold. In the mass region of interest (at 1680\,MeV) the contribution shows no significant feature.

\subsection{\label{Compton}\boldmath $N(1685)$ in Compton scattering }

A trace of $N(1685)$ may have been found in the
$\gamma n\to \gamma n$ total cross section \cite{Kuznetsov:2010as} and in the beam asymmetry 
for $\gamma p\to \gamma p$  \cite{Kuznetsov:2015nla}. We do not see how the two phenomena could possibly be related to the interference pattern in the $\gamma n\to \eta n$ discussed in this paper.

\section{Discussion and Summary}

We have scrutinized the evidence for the existence of a narrow resonance at 1685\,MeV. A structure at this mass was reported from different reactions: i) a very significant bump at this mass in the $\gamma n\to \eta n$ total cross section was observed in three experiments, at GRAAL~\cite{Kuznetsov:2006kt}, ELSA~\cite{Jaegle:2008ux,Jaegle:2011sw}, and MAMI \cite{Werthmuller:2014,Werthmuller:2013rba}; ii) the reaction $\gamma p\to \eta p$ showed a small anomaly at 1685\,MeV~\cite{McNicoll:2010qk}; and iii) an excess of events at about this mass was observed in Compton scattering $\gamma n\to \gamma n$ \cite{Kuznetsov:2010as}. The beam asymmetry for $\gamma p\to \gamma p$  \cite{Kuznetsov:2015nla} reported two structures, one at 1680\,MeV. Intriguingly, the observed properties of the structure - when interpreted as a resonance - agreed very well with predictions of the soliton model for the non-strange member of a $J^P=1/2^+$ antidecuplet of pentaquark states.

i) The new and very precise data from MAMI enabled us to a much more solid partial-wave analysis of the $\gamma n\rightarrow n\eta$ reaction. Our fit results show that the bump in the total cross section and also the behavior of the angular distributions can be understood quantitatively as interference between the two well-known resonances in the $J^P=1/2^-$ wave, the $N(1535)1/2^-$ and the $N(1650)1/2^-$ states. This fit requires, however, that the sign of the electromagnetic $A_{1/2}$ helicity coupling of the $N(1650)1/2^-$ is inverted for the neutron with respect to the current PDG \cite{Beringer:1900zz} entry and also with respect to an early analysis in the framework of the BnGa model \cite{Anisovich:2008wd} (but in agreement with a later BnGa analysis \cite{Anisovich:2013jya}).
 When a narrow $N(1685)1/2^+$ resonance was enforced in the model, the quality of the fit deteriorated significantly. Consequently, there is no evidence for such a state from $\gamma n\rightarrow n\eta$. It is worthwhile to mention that none of the papers reporting evidence for the structure took the interference between the two $J^P=1/2^-$ resonances into account when fitting the data. The $p$ and $n$ helicity couplings of $N(1650)1/2^-$ determined here imply, however, that the $N(1650)1/2^-$ resonance should have a large $s\bar s$ component in its wave function \cite{Boika:2014aha}. This conclusion is avoided when the K-matrix poles are interpreted as ``undressed'' resonance and the K-matrix couplings are confronted with models. The ``dressed'' helicity amplitudes are not in conflict with model calculations.

ii) The angular distributions show that the bump in the total cross section is an S-wave phenomenon; it is not a P-wave enhancement.   

iii) The anomaly at 1685\,MeV in the total cross section of the reaction $\gamma p\to \eta p$ reported in \cite{McNicoll:2010qk} could be traced quantitatively to the opening of the $K\,\Sigma$ threshold. Since data on $\gamma p\to K\,\Sigma$ are included in the Bonn-Gatchina partial wave analysis, there is no free parameter available to fit the shape of the anomaly in the $\gamma p\to \eta p$ cross section. The small size of this anomaly rules out the possibility that the $K\,\Sigma$ threshold might be responsible for the narrow bump observed in the $\gamma n\to \eta n$ total cross section. 
   
iv) The small excess in the number of events in Compton scattering $\gamma n\to \gamma n$ \cite{Kuznetsov:2010as} at 1680\,MeV. A study of the beam asymmetry $\Sigma$ for Compton scattering off  protons reported two narrow structures with masses near 1680 and 1.720\,MeV \cite{Kuznetsov:2015nla}. Their origin needs further clarification.\\[2ex]
 
{\small We would like to thank M. Dieterle for providing the data on $\gamma n \to \pi^0n$ in numerical form. We acknowledge support from the Deutsche Forschungsgemeinschaft (within the SFB/TR16), the Schweizerische
Nationalfonds, and the Russian Foundation for Basic Research.}

\end{document}